\documentclass[aps, pre, preprint, groupedaddress, floatfix]{revtex4-1}

\usepackage{subfigure}
\usepackage{xspace}
\usepackage{nicefrac, amsmath, natbib, amssymb, amsthm, graphicx,tikz}
\usepackage{graphics}
\usepackage{siunitx}
\DeclareMathOperator*{\argmax}{arg\,max}

\newcommand{\sd}{\mathrm{d}}
\newcommand{\e}{\mathrm{e}}

\newcommand{\tm}{\textsuperscript{TM}\xspace}

\RequirePackage{color}
\RequirePackage[colorlinks, urlcolor=my-blue,linkcolor=my-red,citecolor=my-green]{hyperref}
\definecolor{my-blue}{rgb}{0.0,0.0,0.6}
\definecolor{my-red}{rgb}{0.5,0.0,0.0}
\definecolor{my-green}{rgb}{0.0,0.5,0.0}
\definecolor{nicos-red}{rgb}{0.75,0.0,0.0}
\definecolor{light-gray}{gray}{0.6}
\definecolor{really-light-gray}{gray}{0.8}
\definecolor{sussexg}{rgb}{0.0,0.5,0.5}
\definecolor{sussexp}{rgb}{0.5,0.0,0.5}

\theoremstyle{plain}

\theoremstyle{definition}
\newtheorem{definition}{Definition}
\newtheorem{example}{Example}

\theoremstyle{remark}
\newtheorem{remark}{Remark}

\newcommand{\be}{\begin{equation}}
\newcommand{\ee}{\end{equation}}

\usepackage[british]{babel}

\def\m1{\mathbf{1}}

\definecolor{darkgreen}{rgb}{0.0,0.5,0.0}
\definecolor{darkblue}{rgb}{0.0,0.0,0.3}
\definecolor{nicosred}{rgb}{0.65,0.1,0.1}
\definecolor{light-gray}{gray}{0.7}

\begin{document}


\title{\textcolor{black}{Performance of information criteria used for model selection of Hawkes process models of financial data}}

\author{J. M. Chen}
\affiliation{School of Mathematics, Cardiff University, UK}

\author{A. G. Hawkes}
\affiliation{School of Management, Swansea University, UK}

\author{E. Scalas}
\email[]{E.Scalas@sussex.ac.uk}
\affiliation{School of Mathematics and Physical Sciences, University of Sussex, UK}

\author{M. Trinh}
\email[]{M.Trinh@sussex.ac.uk}
\affiliation{School of Mathematics and Physical Sciences, University of Sussex, UK}


\date{\today}

\begin{abstract}
We test three common information criteria (IC) for selecting the order of a Hawkes process with an intensity kernel that can be expressed as a mixture of exponential terms. These processes find application in high-frequency financial data modelling. The information criteria are Akaike's information criterion (AIC), the Bayesian information criterion (BIC) and the Hannan-Quinn criterion (HQ). Since we work with simulated data, we are able to measure the performance of model selection by the success rate of the IC in selecting the model that was used to generate the data. In particular, we are interested in the relation between correct model selection and underlying sample size. The analysis includes realistic sample sizes and parameter sets from recent literature where parameters were estimated using empirical financial intra-day data. We compare our results to theoretical predictions and similar empirical findings on the asymptotic distribution of model selection for consistent and inconsistent IC.
\end{abstract}


\keywords{Hawkes process, self-exciting process, model selection, information criterion, AIC, BIC, HQ}

\maketitle


\section{Introduction}
Technological advancement made it possible to record detailed data of all trades on financial markets. This development called for suitable econometric models that incorporate the time structure of durations between trades. Previously, models were designed such that this information was lost due to aggregation of data to equidistant time grids. However, empirical studies of high-frequency trading data show that intra-day trades have a typical pattern: there is high trading activity at the beginning and end of the trading day whereas there is low trading activity during lunch hours in the middle of the trading day (see for example \cite{Bertram_2004}). Engle and Russell were among the first to propose a point process approach to modelling durations between trades (\cite{Engle_Russell_1997,Engle_Russell_1998}, \cite{Engle_2000}). The proposed autoregressive conditional duration (ACD) model is closely related to the popular GARCH model for volatility clustering. It is also known under the name multiplicative error model (see \cite{Hautsch_2012} for further details on this topic).\\
However, self-exciting point processes have gained vast popularity among econometricians and financial mathematicians. Especially Hawkes processes \citep{Hawkes_1971b, Hawkes_1971a} offered an intuitive notion of endogenous and exogenous components contributing to (trade) event clustering, which is sometimes referred to as ``market reflexivity'' (\cite{Filimonov_Sornette_2012}, \cite{Hardiman_2013}). Additionally, from a theoretical point of view, \cite{Daley_2003} draw the analogy between the role of Hawkes processes spectral approximations of point processes and the importance of autoregressive models for mean square continuous processes.\\
Hawkes processes were originally used for seismic data (\cite{Hawkes_1973}, \cite{Ogata_1988}), but their characteristic property of self-excitation and event clustering are appealing properties for mimicking similar phenomena found as stylized facts in intra-day financial data. \cite{Bowsher_2007} was among the early works to establish the connection between Hawkes processes and financial modelling. As there is an intensity-based as well as cluster-based definition, there exist various simulation and estimation techniques which take advantage of either perspective on Hawkes processes. To mention a few, for simulation we have the thinning approach \citep{Ogata_1981}, the time-change approach based on the random time-change theorem \citep{Meyer_1971} and applied specifically to Hawkes processes for instance in \citep{Ozaki_1979}, exact simulation \citep{Dassios_Zhao_2013} and perfect simulation \citep{Moller_Rasmussen_2005}. Concerning estimation techniques, the standard maximum likelihood approach can for example be found in \cite{Ozaki_1979}. Beyond that, \cite{Hawkes_1973} used a spectral estimation approach,  \cite{Rasmussen_2013} proposes a Bayesian estimation technique and an application of the expectation maximization (EM) algorithm can be found in \cite{Veen_Schoenberg_2008}.\\
These tools for handling Hawkes processes numerically paved the way for applications on various types of financial data such as mid-price changes, order books, extreme price movements (among others) gathered from liquid stocks, futures, indices or foreign exchange markets. For details, the review paper by \cite{Bacry_2015} gives a very good summary of recent literature on Hawkes models in finance.\\
Essentially, for parametric estimation, there are two kernels which are widely used in the literature to fit financial data: the exponential kernel and the power law kernel. Whereas the power law asymptotics are additionally backed up by results from non-parametric estimation literature as in \cite{Bacry_2012}, the exponential kernel case is analytically more tractable and is still applied in recent literature (\cite{Hardiman_2013}, \cite{Rambaldi_2015}, \cite{Lallouache_Challet_2016}).\\
Today's computing power not only allows accurate recording of high-frequency trades, but enables us to fit almost arbitrarily complex models to previously gathered data. Recent proposals to model such data include intensities of Hawkes processes that can be expressed as weighted sums of exponential and power law kernels. The natural question arises as to how many terms should be included in such a model to be best suited in describing the data. Information criteria (IC) offer quantitative methods to discriminate between (possibly numerous) models. There are two competing objectives when it comes to selecting an ``optimal'' model order: On the one hand we would like to capture and describe the observed phenomena within the data as accurately as possible but, on the other hand, it is important to keep the complexity of the model to a minimum. A complex model can lead to numerical instabilities and superfluous parameters that do not carry much descriptive power. Information criteria are quantitative tools to manage this trade-off situation. Our aim in this paper is to test how well this model selection method could work for a Hawkes process intensity of weighted sums of exponential terms using simulated data.\\
The paper is organized as follows: Section \ref{sec:hawkes-model-wit} is devoted to the exponential Hawkes-P model. After a short definition and discussion of the average intensity, we move on to the simulation procedure and the parameter estimation method via maximum likelihood. In Section  \ref{sec:model-selection}, we give a short introduction to information criteria and discuss the consistency property. Finally, we describe the setup of the Monte-Carlo experiment and give the numerical results in Section~\ref{sec:numerical-results-1}.

\section{A Hawkes model with exponential kernels}
\label{sec:hawkes-model-wit}
For a self-exciting point process $(N(t))_{t\geq 0}$, the conditional intensity function is formally defined by

\begin{equation}
  \label{eq:34}
  \lambda(t|\mathcal{F}_t) := \lim_{\Delta t \to 0} \mathbb{P} (N(t+\Delta t) - N(t) =1| \mathcal{F}_t)/\Delta t,
\end{equation}
where $\mathcal{F}_t$ represents the known history up to time $t$.
We assume the conditional intensity function to be of the form (conditioning on history removed for the sake of simplicity)
\begin{equation}
  \lambda(t) = \mu + \int_0^t g(t-\tau) \,\sd N(\tau), \label{eq:2}
\end{equation}
where we have for the response function $g(\tau) \geq 0$ $\forall \tau \in \mathbb{R}^{+}$ and $\mu > 0 $ is the baseline intensity. The term containing the response function can be identified with the self-excitation property and is therefore referred to as the endogenous part of the intensity whereas the baseline intensity is the exogenous part.
The above intensity function defines a Hawkes process with finite past, as we assume the counting process $(N(t))_{t\geq 0}$ to start at $0$. Note that we deviate from the original definition, where usually the integral in (\ref{eq:2}) is evaluated over $(-\infty, t]$.\\
In particular, we are interested in the case when the response function can be written as a weighted sum of exponentials:
\begin{equation}
  \label{eq:g}
  g(t) = \sum_{m=1}^P \alpha_m \e^{-\beta_m t}.
\end{equation}
Then, the intensity function is given by
\begin{equation}
  \label{eq:1}
  \lambda(t) = \mu + \sum_{m=1}^P \alpha_m \sum_{i=1}^k\e^{-\beta_m (t-t_i)}
\end{equation}
with $\mu, \alpha_m, \beta_m > 0$ and $\{t_1, \ldots, t_k\}$ are the jump times of $N(t)$ up to time $t$. In short, we will call this process exponential Hawkes-$P$ process, where $P$ is the order of the process.\\
We will consider this class of Hawkes processes as a possible parametric model for durations between trades.

\paragraph{Average intensity: stationary vs. non-stationary case}
In \cite{Hawkes_1971a} the average intensity for a stationary Hawkes process with infinite past has been calculated to be
\begin{equation}
  \label{eq:4}
  \Lambda := \mathbb{E}[\lambda(t)] = \frac{\mu}{1 - \int_0^\infty g(\nu) \,\sd \nu},
\end{equation}
where $n := \int_0^\infty g(\nu) \, \sd \nu$ is called the branching ratio. This result follows essentially by taking the expectation on both sides of (\ref{eq:2}). In particular, for the exponential kernel we have $n= \sum_{m=1}^P \frac{\alpha_m}{\beta_m}$ and the stationarity condition is $n < 1$. The special case of $n=1$ also allows stationary processes which are treated in \cite{Bremaud_Massoulie_2001}.\\

However, for the Hawkes process with finite past associated with the intensity in (\ref{eq:1}), we apply a different approach using Laplace transforms: Let $\varphi(t) := \mathbb{E}[\lambda(t)]$ now be the average intensity function of a non-stationary Hawkes process. Then, taking expectations on both sides of (\ref{eq:2}) yields
\begin{equation}
  \label{eq:9}
  \varphi(t) = \mu + \sum_{m=1}^P \int_0^t \alpha_m \e^{-\beta_m(t-u)}\varphi(u) \,\sd u.
\end{equation}

The Laplace transform of $\varphi$ is given by
\begin{align}
  \tilde{\varphi}(s) &= \int_0^\infty \e^{-st} \varphi(t) \,\sd t \nonumber\\
  &= \int_0^\infty \e^{-st} \mu \, \sd t + \sum_{m=1}^P \alpha_m \int_{t=0}^{\infty} \e^{-st} \int_{u=0}^t \e^{-\beta_m(t-u)} \varphi(u) \, \sd u \,\sd t \nonumber\\
  &= \frac{\mu}{s} + \sum_{m=1}^P \alpha_m \int_{u=0}^\infty \e^{-su}  \varphi(u) \int_{t=u}^\infty \e^{-(s+\beta_m)(t-u)} \, \sd t \, \sd u \label{eq:5}\\
  &= \frac{\mu}{s} + \sum_{m=1}^P \frac{\alpha_m}{s+\beta_m} \int_{u=0}^\infty \e^{-su}  \varphi(u) \, \sd u = \frac{\mu}{s} + \left(\sum_{m=1}^P \frac{\alpha_m}{s+\beta_m}\right) \tilde{\varphi}(s), \nonumber
\end{align}
where in (\ref{eq:5}) we we are able to apply Fubini's theorem since the integrand is positive. Finally, we have an algebraic equation which can be solved for $\tilde{\varphi}$:
\begin{equation}
  \tilde{\varphi}(s) = \frac{\frac{\mu}{s}}{1 - \sum_{m=1}^P \frac{\alpha_m}{s + \beta_m}}\label{eq:14}.
\end{equation}
For $P>1$ we could write alternatively:
\begin{equation}
  \tilde{\varphi}(s) = \frac{\mu}{s} \frac{\prod_{m=1}^P (s + \beta_m)}{\prod_{m=1}^P(s+\beta_m) - \sum_{m=1}^P \alpha_m \prod_{k\neq m}(s+\beta_k)}. \label{eq:10}
\end{equation}
This gives an analytic expression for the Laplace transform of the intensity function. From Equation (\ref{eq:14}) we can see that it is reasonable to demand the usual stationarity condition $\sum_{m=1}^P \frac{\alpha_m}{\beta_m} < 1$ in order to ensure that the right hand side term is well defined.

In general, the evaluation of the average intensity function can be done by (numerical) Laplace inversion. However, for lower model orders (up to $P=4$) it is possible to invert the Laplace transform analytically. We will show this for first and second order in the following examples.

\begin{example}[Formula for the average intensity in the case $P=1$]
  For $P=1$ the expression in (\ref{eq:14}) simplifies to
  \begin{equation}
    \label{eq:11}
    \tilde{\varphi}(s) = \frac{\mu(s+\beta_1)}{s(s+\beta_1-\alpha_1)} = \frac{\mu}{\beta_1-\alpha_1}\left( \frac{\beta_1}{s} - \frac{\alpha_1}{s+\beta_1-\alpha_1}\right),
  \end{equation}
where we used a partial fractions decomposition in the last step. This allows us to analytically invert the Laplace transform:
\begin{equation}
  \label{eq:12}
  \varphi(t) = \frac{\mu}{\beta_1-\alpha_1}\left( \beta_1 - \alpha_1\e^{-(\beta_1-\alpha_1)t}\right), \quad t>0.
\end{equation}
\end{example}

\begin{figure}
  \centering
  \includegraphics[scale=0.5]{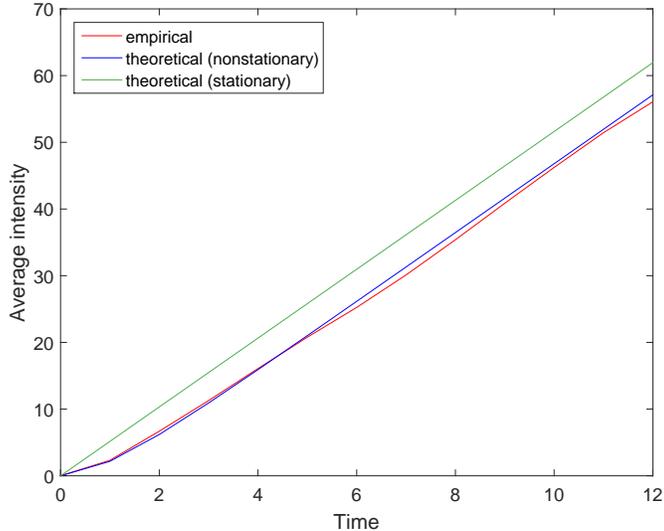}
  \caption{A comparison between the average number of events from a MC simulation and the theoretical values. For the parameter values $\mu=0.5$, $\alpha_1=3.1, \alpha_2=5.9, \beta_1=9.9$ and $\beta_2=10$, we simulated an exponential Hawkes process of order $P=2$ with finite past and plotted the empirical average number of events (red curve) against the theoretical values of the expected number of events in Eq (\ref{eq:19}). In the non-stationary case, we integrate the average intensity function in Eq (\ref{eq:17}) which corresponds to the blue curve. The stationary case is shown via the green curve.}
  \label{fig:testMeanIntensity}
\end{figure}

\begin{example}[Formula for the average intensity in the case $P=2$]
  For $P=2$ we have
  \begin{equation}
    \label{eq:15}
    \tilde{\varphi}(s) = \frac{\mu (s + \beta_1)(s +\beta_2)}{s[(s+\beta_1)(s+\beta_2) - \alpha_1(s+\beta_2) - \alpha_2(s+\beta_1)]}
  \end{equation}
  Starting from order $P=2$, the explicit formulas can be quite complicated. 


Let $R$ and $Q$ denote the polynomial in the numerator and the denominator of the right hand side expression in (\ref{eq:15}) respectively. Then, assuming $Q$ has only real valued roots of single multiplicity denoted by $s_1, s_2, s_3$, the partial fractions decomposition is given by
\begin{equation}
  \label{eq:16}
  \tilde{\varphi}(s) = \frac{P(s)}{Q(s)} = \sum_{i=1}^3 \frac{P(s_i)}{Q'(s_i)(s-s_i)} = \mu \left( \frac{A_1}{s} + \frac{A_2}{s-s_2} + \frac{A_3}{s-s_3}\right), 
\end{equation}
where

\begin{gather}
  s_1 = 0, \quad s_2 = \frac{1}{2}(\gamma - \xi), \quad s_3 = \frac{1}{2}(\gamma + \xi) \label{eq:25}\\
  \text{with} \quad \gamma = \alpha_1 + \alpha_2 - \beta_1 - \beta_2 \quad \text{and} \quad
  \xi = \sqrt{\gamma^2 - 4(\beta_1\beta_2 - \alpha_1\beta_2 - \alpha_2\beta_1)}.
\end{gather}

The partial fractions decomposition implies that
\begin{equation}
  \label{eq:24}
  A_1(s-s_2)(s-s_3) + A_2s(s-s_3) + A_3s(s-s_2) \stackrel{!}{=} (s+\beta_1)(s+\beta_2)
\end{equation}
and comparing coefficients of $s^2, s$ and $1$ on both sides of the equation yields
\begin{align}
  A_1 + A_2 + A_3 &= 1 \label{eq:28}\\
  -A_1(s_1+s_2) - A_2s_3 - A_3s_2 &= \beta_1 + \beta_2 \label{eq:29}\\
  A_1s_1s_2 &= \beta_1\beta_2. \label{eq:30}
\end{align}
Then we get
\begin{equation}
  \label{eq:27}
  A_1 = \frac{\beta_1\beta_2}{s_1s_2} = \frac{\beta_1\beta_2}{(\gamma^2 - \xi^2)/4} = \frac{\beta_1\beta_2}{\beta_1\beta_2-\alpha_1\beta_2-\alpha_2\beta_1}
\end{equation}
by solving (\ref{eq:30}) for $A_1$ and inserting (\ref{eq:25}).\\
Now multiply (\ref{eq:28}) by $s_2$ and add (\ref{eq:29}) to get
\begin{equation}
  \label{eq:31}
  -A_1s_3 + A_2(s_2-s_3) = \beta_1 + \beta_2 + s_2.
\end{equation}
Solving for $A_2$ we get
\begin{align}
 A_2 &= \frac{\beta_1\beta_2/s_2 + \beta_1 + \beta_2 + s_2}{s_2-s_3} 
  = \frac{\beta_1\beta_2 + s_2(\beta_1+\beta_2) + s_2^2}{s_2(s_2-s_3)} \nonumber\\
  &= \frac{4\beta_1\beta_2 - 2(\xi-\gamma)(\beta_1+\beta_2) + (\xi-\gamma)^2}{2\xi(\xi-\gamma)}
  = \frac{(\xi-\gamma-2\beta_2)(\xi-\gamma-2\beta_1)}{2\xi(\xi-\gamma)} \nonumber\\
  &= \frac{(\xi - \alpha_1 -\alpha_2 +\beta_1 -\beta_2)(\xi -\alpha_1 -\alpha_2 -\beta_1 +\beta_2)}{2\xi(\xi-\gamma)}.
\end{align}
Multiplying (\ref{eq:28}) by $s_3$, adding (\ref{eq:29}) and following similar steps as for $A_2$ yield
\begin{equation}
  \label{eq:32}
  A_3 = \frac{(\xi+\gamma+2\beta_1)(\xi+\gamma+2\beta_2)}{2\xi(\xi+\gamma)}
  = \frac{(\xi+\alpha_1+\alpha_2+\beta_1-\beta_2)(\xi+\alpha_1+\alpha_2-\beta_1+\beta_2)}{2\xi(\xi+\gamma)}.
\end{equation}
The Laplace inversion gives
\begin{equation}
  \label{eq:17}
  \varphi(t) = \mu\left( A_1 + A_2 \e^{s_2 t} + A_3 \e^{s_3 t}\right).
\end{equation}
Note that with the condition $\sum_{m=1}^P \frac{\alpha_m}{\beta_m} < 1$ it follows that the roots $s_2$ and $s_3$ are real and negative.
\end{example}
From both examples, we can see that for large times $t$ the exponential terms in Equations (\ref{eq:12}) and (\ref{eq:17}) become negligible and the remaining expressions coincide with the intensity function of the stationary case. In a small Monte-Carlo (MC) experiment, we simulated 1000 paths of a Hawkes process with 1000 events (see also \texttt{empirAgg2.m}). The parameters are $\mu=0.5$, $\alpha_1=3.1$, $\alpha_2=5.9$, $\beta_1=9.9$ and $\beta_2=10$. Figure \ref{fig:testMeanIntensity} shows a plot of the empirically observed average number of events against the theoretically expected number of events. Plotting such figures might be useful for validation of a simulations algorithm. Recall the relation between average intensity function $\varphi$ and expected number of events of a point process $(N(t))_{t\geq 0}$:
\begin{equation}
  \label{eq:19}
  \mathbb{E}[N(t)] = \int_0^t \varphi(\tau) \, \sd \tau
\end{equation}
For small times we can observe the transient exponential behavior which
vanishes for large times. In particular, the slope of the two theoretical functions are approximately equal for large times and indicate that the intensity function of the non-stationary case converges to the stationary case. Also, we can verify the edge effect when simulating a Hawkes process with finite past, which will be briefly discussed in the next section.

\subsection{Simulation}
\label{sec:simulation}
As seen in the previous section, simulating a Hawkes process with finite past in order to approximate a Hawkes process with infinite past will cause the simulated process to be non-stationary at the beginning of the simulation time. This phenomenon is also known as edge effect as offspring of events that might have occurred in the past are omitted. For further details on this see \cite{Moller_Rasmussen_2005, Moller_Rasmussen_2006}.\\
However, similar to \cite{Dassios_Zhao_2013}, we explicitly want to work with a Hawkes process with finite past. Therefore, we view the edge effect as an inherent property of the model rather than an artifact of the simulation. Besides, the exact simulation algorithm in \cite{Dassios_Zhao_2013}, though applicable to multidimensional exponential models, does not directly apply to our proposed model due to the lack of identification of the exogenous and endogenous part of the intensity. This leaves us with the popular thinning algorithm going back to \cite{Lewis_Shedler_1979} and \cite{Ogata_1981}. We used an implementation of the thinning algorithm to simulate the process on a time interval $[0,T]$ (see \texttt{hawkesThinning.m}) and compare models up to order $3$. We first generate sample data that serve as a technical example for the estimation and model selection methods. The parameter settings are given in Table \ref{hawkesSim}.

\paragraph{Connection to empirical findings in financial literature}
In order to enhance the practical relevance of our experiments and results we would also like to use parameter settings which allow intensities which can also be observed in empirical studies.
Concerning the exponential Hawkes-P model, \cite{Hardiman_2013} found that the use of the single exponential intensity function might give misleading results, which is also confirmed by \cite{Rambaldi_2015}. However, this does not necessarily hold for exponential Hawkes processes of higher order: \cite{Lallouache_Challet_2016} found that Hawkes models with exponential intensity kernels of order $P$ greater than one perform better than the single exponential model and comparably well to power law models when applied to FX data. This is why we include a parameter set that was estimated in this paper for our MC experiment (see Table \ref{param_lallouache}).

\subsection{Maximum likelihood estimation (MLE) and goodness of fit}
\label{sec:mle-goodness-fit}
Although we are primarily interested in the performance of model selection, we must make sure that the MLE gives reasonable results. This is because we expect a close connection between the quality of the MLE and the subsequent model selection result. A poor MLE due to numerical problems or lack of data is likely to compromise the model selection. For example, a correctly selected model order can be meaningless if the estimated model itself fails to describe and predict key features or quantities of the data we are interested in. In the following subsections we briefly present the fitting procedure as well as the root mean squared error as our chosen measure for goodness of fit. 

\subsubsection{Fitting via MLE}
\label{sec:fitting}
The fitting algorithm follows the theory in \cite{Ozaki_1979} which is a standard maximum likelihood procedure. For a self-exciting point process with intensity $\lambda$ the log-likelihood for data $0< t_1 < \ldots < t_n < T$ is given by
\begin{equation}
  \label{eq:26}
  \log\mathcal{L}(t_1, \ldots, t_n \vert \theta) = - \int_0^T \lambda(t\vert \theta) \,\sd t + \int_0^T \log(\lambda(t|\theta)) \sd N(t).
\end{equation}
Let $\theta=(\mu, \alpha_1, \ldots, \alpha_P, \beta_1. \ldots, \beta_P)$ be the vector of parameters for the Hawkes-$P$ model. Inserting Eq (\ref{eq:1}) into Eq (\ref{eq:26}) gives
\begin{align}
  \label{eq:18}
 \log\mathcal{L}(t_1, \ldots, t_n \vert \theta) = -\mu T &- \sum_{m=1}^P \left[\frac{\alpha_m}{\beta_m} \sum_{t_i < T} \left( 1 - \e^{-\beta_m(T-t_i)}\right)\right] \nonumber\\
  &+ \sum_{t_k < T} \log\left( \mu + \sum_{m=1}^P \alpha_m \sum_{t_i < t_k} \e^{-\beta_m(t_k-t_i)}\right).
\end{align}
Moreover, \cite{Ozaki_1979} shows that the log-likelihood can be calculated recursively, which reduces the computational burden from $\mathcal{O}(n^2)$ to $\mathcal{O}(n)$: Assume that $T=t_n$, i.e. the last event is the last time point of observation. Then
\begin{align}
  \log\mathcal{L}(t_1, \ldots, t_n \vert \theta) &= -\mu t_n - \sum_{m=1}^P \left[\frac{\alpha_m}{\beta_m} \sum_{t_i \leq t_n}\left( 1 - \e^{-\beta_m(t_n-t_i)}\right)\right]\nonumber\\
 &\hphantom{{}= -\mu t_n}{}+ \sum_{t_k \leq t_n} \log\left( \mu + \sum_{m=1}^P \alpha_m A_m(k)\right),\\
 \text{ where } A_m(1) &= 0 \quad \forall\, m = 1, \ldots, P\nonumber\\
A_m(k) &= \sum_{t_i < t_k} \e^{-\beta_m(t_k-t_i)} = \left(1 + A_m(k-1)\right) \e^{-\beta_m(t_k-t_{k-1})}.\nonumber
\end{align}
To obtain the MLE of the parameters we maximize the log-likelihood function with respect to the parameters subject to the stationarity condition:
\begin{gather} 
  \argmax_{\mu, \alpha_1, \ldots, \alpha_P, \beta_1, \ldots, \beta_P} \log\mathcal{L}(t_1,\ldots, t_n \vert \mu, \alpha_1, \ldots, \alpha_P, \beta_1, \ldots, \beta_P)\\
\text{s.t. } \mu, \alpha_1, \ldots, \alpha_P, \beta_1, \ldots, \beta_P > 0,  \quad \beta_1 < \ldots < \beta_P 
\quad \text{ and }\quad \sum_{m=1}^P \frac{\alpha_m}{\beta_m} < 1.\nonumber
\end{gather}
We assume the $\beta$ parameters to be ordered to avoid identification problems. The maximization (or rather the minimization of the negative log-likelihood) is typically done numerically as the estimators are not available in closed form. We used the standard MATLAB\tm function \texttt{fmincon} for constrained problems. The optimization routine can be found in the supplementary files \texttt{fitting.m}, \texttt{conditions.m} and \texttt{LogLik\_iter.m}.

\subsubsection{Goodness of fit}
\label{sec:goodness-fit}
Important asymptotic properties of the MLE for Hawkes processes have been studied and proven by \cite{Ogata_1978} (see also in the appendix in \cite{Rambaldi_2015} for a brief summary). In particular, we may assume the MLE to be consistent, i.e. with sample size tending to infinity the MLE converge to the true values of the parameters. In order to verify these results with our MC experiment, we use the RMSE (root mean squared error) as a measure for the goodness of fit: Let $\theta$ be a generic model parameter to be estimated and $\hat{\theta}$ the corresponding estimator. We are given $N=1000$ samples and have the parameter estimates $\hat{\theta}^{(k)}$, $k=1,\ldots,N$. For each sample we calculate the (absolute) root mean squared error to be
\begin{equation}
  \label{eq:13}
  \text{RMSE}(\theta) =  \sqrt{\frac{1}{N} \sum_{k=1}^N \vert\theta - \hat{\theta}^{(k)}\vert^2}
\end{equation}
and the relative root mean squared error
\begin{equation}
  \label{eq:3}
  \text{RMSE}_{\text{rel}}(\theta) = \frac{1}{\theta}\sqrt{\frac{1}{N} \sum_{k=1}^N \vert\theta - \hat{\theta}^{(k)}\vert^2}.
\end{equation}
It is easy to calculate the above quantities as the true model values are known in our mock data setting.


\section{Information criteria and model selection}
\label{sec:model-selection}

In the following sections we will define the IC we are interested in and briefly describe relevant theoretical concepts. Based on that, we present the results of a simple MC experiment to assess model selection using IC. We are aware that certain conditions of our experiment are not given in reality and therefore also discuss the limitations to the conclusions we may draw from the numerical results.

\paragraph{Definitions and theoretical properties}
\label{sec:defin-theor-prop}

This section follows introductory work which can be found in \cite{Claeskens_Hjort_2008} and references therein.

\begin{definition}
  For a given model fitted to data via MLE let $\mathcal{L}$ be the maximal log-likelihood value, $k$ the number of parameters and $n$
  be the sample size of the data set. Then we define:
  \begin{enumerate}
  \item \textbf{Akaike's information criterion (AIC)} (see
    \cite{Akaike_1973})
    \begin{equation}
      \label{eq:6}
      \text{AIC} = -2\mathcal{L} + 2k
    \end{equation}
  \item \textbf{Bayesian information criterion (BIC)} (see
    \cite{Schwarz_1978})
    \begin{equation}
      \label{eq:7}
      \text{BIC} = -2\mathcal{L} + k\ln(n)
    \end{equation}
  \item \textbf{Hannan and Quinn information criterion (HQ)} (see
    \cite{Hannan_1979} and \cite{Hannan_1980})
    \begin{equation}
      \label{eq:8}
      \text{HQ} = -2\mathcal{L} + 2k \ln(\ln(n)) 
    \end{equation}
  \end{enumerate}
\end{definition}
The exponential Hawkes-P process from Section \ref{sec:hawkes-model-wit} is an example of a nested series of models with $1+2P$ parameters and are therefore quite suitable for calculating information criteria. The formulas for the IC were implemented in the function \texttt{IC.m}. The above information criteria are of the form
\begin{equation}
  \label{eq:20}
  \text{IC}(M_k) = -2\mathcal{L} + c(k,n)
\end{equation}
where $M_k$ is a model associated with parameter number $k$ and $c(k,n)$ is a suitably chosen penalty term that accounts for the complexity of the model, i.e. the number of  parameters. Within a given set of models to choose from, the ``best'' model is the one which minimizes the IC value. In other words, the selected model should give the best fit to the data, i.e. have a large log-likelihood value, while being as parsimonious as possible, i.e. use few parameters. Therefore, formula (\ref{eq:20}) represents the trade-off situation we have discussed previously.
\begin{remark}\hfill
\label{rem:defin-theor-prop-1}
  \begin{enumerate}
  \item The AIC was derived from estimating the Kullback-Leibler distance between the ``true'' model distribution and the estimated one. \cite{Hurvich_Tsai_1989} proposed a correction of the AIC for small samples:
    \begin{equation}
      \label{eq:33}
      \text{AICc} = -2\mathcal{L} + \frac{2kn}{n-k-1}.
    \end{equation}
We shall follow the recommendation in \cite{Burnham_Anderson_2004} and use the AICc whenever $n < 40 k_{\text{max}}$ as a rule of thumb, where $k_{\text{max}}$ is the maximal number of parameters used among the candidate models.
  \item The BIC was first derived in a Bayesian estimation approach, but is also valid in the frequentist context and there is an alternative derivation of the BIC from the frequentist perspective (see \cite{Burnham_Anderson_2004} for details)
  \item The HQ is designed to have the slowest growing penalty term that still renders the IC to be strongly consistent (see later for a more precise definition). The proof makes use of the law of iterated logarithm. Besides, the HQ was originally defined more generally as 
    \begin{equation}
      \label{eq:21}
      \text{HQ}' = -2\mathcal{L} + 2ck\ln(\ln(n)), \quad c > 1,
    \end{equation}
but $c$ was chosen to be $1$ in a subsequent example. \cite{Claeskens_Hjort_2008} point out that the choice of $c$ is not clear and renders the information criterion less relevant for practitioners.
  \end{enumerate}
\end{remark}

Similar to the consistency property of the MLE, it is a desirable property to have the IC selecting the correct model order with high probability when the underlying sample size increases. To be more precise:

\begin{definition}
  Let $n$ be the underlying sample size,  $\mathcal{J}$ be the set of models among all competing models that minimize the Kullback-Leibler distance to the true model and let $\mathcal{J}_0 \subset \mathcal{J}$ be the subset of models with minimal (parameter) dimension. Then, an IC is said to be \textit{consistent} if there is a $j_0 \in \mathcal{J}_0$ such that
  \begin{equation}
    \label{eq:22}
    \lim_{n\rightarrow \infty} \mathbb{P}\left\{ \min_{l \in \mathcal{J}\setminus \mathcal{J}_0}(\text{IC}(M_{j_0})-\text{IC}(M_{l})) > 0 \right\} = 1,
  \end{equation}
i.e. the probability that the IC will choose a model with smallest dimension minimizing the Kullback-Leibler distance converges to 1.\\
An IC is \textit{strongly consistent} if the assertion in (\ref{eq:22}) holds almost surely:
\begin{equation}
  \label{eq:23}
  \mathbb{P}\left\{\min_{l \in \mathcal{J}\setminus \mathcal{J}_0}(\text{IC}(M_{j_0})-\text{IC}(M_{l})) > 0, \quad \text{for almost all } n \right\} = 1
\end{equation}
\end{definition}

\begin{remark}
The above definition follows the notation in \cite[p.~101]{Claeskens_Hjort_2008}, but the original proof of sufficient conditions for consistency and strong consistency are shown in \cite{Sin_White_1996}, (where consistency actually goes under the name of weak consistency).
\end{remark}

As a matter of fact, the AIC fails to be consistent as the penalty term does not depend on the sample size. The asymptotic distribution of the associated model selection was analyzed for autoregressive models for example in \cite{Shibata_1976}. BIC and HQ on the other hand are found to be strongly consistent. As a consequence, their asymptotic distribution of model selection is bound to converge to a delta on the most parsimonious Kullback-Leibler minimizing model. The respective convergence rates for AIC and BIC were analyzed in \cite{Zhang_1993} for another regression model.

\paragraph{Consistency from a practical perspective}
From the previous section one might conclude that the non-consistent AIC would be inferior to the consistent BIC and HQ. However, the situation is more complicated: We have to keep in mind that consistency is an asymptotic property. This means that in theory the consistent IC will eventually outperform the AIC for almost all cases if the sample size is sufficiently large. Unfortunately, practitioners just have a limited amount of data available and it is very difficult to judge whether the sample size belongs to the asymptotic region. Indeed, empirical studies suggest for various statistical models that the AIC outperform the BIC in small sample cases\footnote{By ``small samples'' we refer to the situation that the sample size is not sufficiently large enough for the asymptotic consistency results to hold, but large enough such that effects similar to the paradox discussed by \cite{Freedman_1983} can be safely excluded.}: As an example among regression models, \cite{Hurvich_Tsai_1989, Hurvich_Tsai_1990} compared different IC on simulated data especially to promote the (still inconsistent) AICc as a modification of the AIC for smaller samples. More recently, \cite{Javed_Mantalos_2013} applied IC (AIC, BIC, HQ, AICc) in a MC simulation of (nonlinear) GARCH models. Their results suggest that the AIC outperforms the BIC and HQ for higher-order GARCH processes.\\
As a consequence of the above discussion, we can make the idea and objective of our MC experiment more precise: First, we need to point out that the numerical results of the simplistic setting of our MC experiment do not directly translate to how empirical data should be handled. IC are one of many tools for model-selection and cross-validation. We do not expect to find a ``best'' IC, but rather want to verify the theoretical properties of the different IC for Hawkes processes. In particular, due to the fact that most theoretical results have been derived for regression models only, our work may help to shed light on asymptotic regions and convergence rates of consistent IC and the asymptotic distribution of selected orders of the AIC for this model class. The verification of theoretical properties will be the main aim for the MC simulation using Parameter Set 1 whereas for our empirical Parameter Set 2 we following the advice given in \cite{Burnham_Anderson_2004} and  use realistic sample sizes.

\section{Numerical results}
\label{sec:numerical-results-1}
Using the thinning algorithm described in Section \ref{sec:simulation} we simulated four different data sets containing 1000 samples. Three of them correspond to each row of Parameter Set 1 in Table \ref{hawkesSim} and one data set consists of samples of an exponential Hawkes-2 model with parameter values from Parameter Set 2 shown in Table \ref{param_lallouache}. Especially for Parameter Set 2 the time horizon $T$ can be assumed to be given in seconds. It ranges from \SI[mode=text]{10}{\minute} to \SI[mode=text]{6}{\hour} to reflect typical intra-day financial data sets.\\
In order to check how well the estimation method works for our parameter sets, we first assume that the correct model order $P$ is known and run a MLE of the parameters of the true model underlying each data set. Subsequently, we are able to calculate the RMSE as a measure of distance between the true and the estimated parameter values. The absolute and relative RMSE values for Parameter Set 1 can be found in Table \ref{rmse} and Table \ref{rmseRel} respectively. For Parameter Set 2, see Tables \ref{rmselal} and \ref{rmseRellal}. We observe that the RMSE decreases with increasing sample size. This is to be expected as the MLE is known to be consistent.\\
Finally, we assume that the true model order is not known, but needs to be selected by the IC. Consequently, for each data set we have to fit all possible model orders $P=1,2,3$ and to calculate the associated IC values. In the following we discuss the results of the model selection.\\

We first consider Parameter Set 1. For simulated data with model order $P=1$ we can see in Table \ref{rmseRel} that the relative RMSE is comparably low even for the smallest samples corresponding to the time horizon $T=500$. The model selection in Table \ref{modelSelection01} confirms that the smallest sample size might already be enough to guarantee high success rates (over $90\%$) of all IC. Nevertheless, already in the lowest order case, we can observe the different behavior of consistent and inconsistent IC. For BIC and HQ, the success rate improves with increasing average sample size. In particular, the relation seems to be monotone and, in the case of the BIC, the success rate reaches $100\%$ already for $T=1000$, The HQ performs slightly worse than BIC, but is still well over $90\%$ and very close to $100\%$ for $T=5000$. However, the AIC behaves in a more concerning manner. Even for large sample sizes the model selection using the AIC allows a comparably large probability ($>6\%$) to select a higher order than $P=1$. As the AIC is not a consistent IC, we cannot exclude the possibility that these results already approximate the asymptotic distribution of model selection of the AIC. As mentioned earlier this asymptotic distribution is typically different from the delta distribution with mass one on the true model order. Additionally, the numerical results show that increasing the average sample size does not necessarily increase the success rate of model selection. For instance, moving from $T=500$ to $T=1000$ we can observe a decrease in success rate in the AIC case.\\
In the case of model order $P=2$, there is the possibility of both over- and underestimation. We observe quite large RMSE for the parameters $\alpha_1$ and $\beta_1$, especially for smaller samples corresponding to $T=500$ and $T=1000$. This could be one of the factors affecting the model selection for $T=500$ in Table \ref{modelSelection02}: there is a significant proportion of underestimation among all IC, most notably the high underestimation rate of almost $95\%$ of the BIC. The AIC seems to perform best in this setting for $T=500$ with success rates slightly above $50\%$, but also with $48\%$ underestimation. For larger samples, the BIC and HQ select the correct model order with very high probability (around $90\%$ or even larger) and the BIC reaches $100\%$ success rate at $T=2000$. Again, we have the adverse effect that the success rates of the AIC decrease with growing average sample size. Even for the largest average sample size for $T=5000$ there is a relatively large probability of overestimation of over $6\%$.\\
For data simulated with $P=3$ we have a similar behavior as with $P=2$. Again, Table \ref{rmseRel} reports large RMSE for the parameters $\alpha_1$ and $\beta_1$ in small sample cases. As $P=3$ is the highest selectable model order, this excludes cases of overspecification. This means that the we can observe the same pattern in model selection of the AIC as for the BIC and HQ: Starting at $T=500$, there are mostly cases of underestimates followed by improving success rates as the sample size increases. All IC reach $100\%$ success rate for $T=2000$. However, it is very likely that we would be able to observe the tendency of the AIC to overestimate if we included higher orders $P>3$ in the model selection set.\\

When working with Paramerter Set 2, we chose the time horizons \SI[mode=text]{10}{\minute}, \SI[mode=text]{15}{\minute}, \SI[mode=text]{30}{\minute}, \SI[mode=text]{1}{\hour}, \SI[mode=text]{3}{\hour} and \SI[mode=text]{6}{\hour}. At first, there are large RMSE values for $T=600$ and $T=900$ (see Tables \ref{rmselal} and \ref{rmseRellal}), which shows that the sample sizes are so small that we cannot ensure good estimates of the MLE method. Especially estimates of $\alpha_2$ and $\beta_2$ have large RMSE. This situation corresponds to the case $T=500$ in the setting of Parameter Set 1. When we compare with the corresponding model selection in Table \ref{modelSelectionlallouache}, we observe the same phenomenon of underestimation is most severe for the BIC, less for the HQ and least for the AIC. As samples are quite small for these cases and may fulfill the rule of thumb discussed in point (i) in Remark \ref{rem:defin-theor-prop-1}, we included the combined model selection rule AICc/AIC in the table. It applies the AICc whenever $n < 3\cdot 40 = 120$ and the AIC otherwise. The numerical results for the combined AICc/AIC selection rule are very similar to the standalone AIC and even slightly worse for T=600 and T=900.\\
When we move on to larger samples from \SI[mode=text]{30}{\minute} to \SI[mode=text]{1}{\hour}, there is a noticeable change in the RMSE values. More precisely, the RMSE values decrease faster for the second exponential term, i.e. $\alpha_2$ and $\beta_2$, which leads to the first exponential term with $\alpha_1$ and $\beta_1$ to contribute more to the overall estimation error. There is a noticeable increase in the rate of correct model selection among all IC ranging over $90\%$ for $T=3600$.\\
Finally, for large samples with time horizons from \SI[mode=text]{3}{\hour} up to \SI[mode=text]{6}{\hour} represent data of half up to an entire trading day respectively. The relative RMSE of each parameter is less than $20\%$ and the rate of correct model selection for the consistent IC (BIC and HQ) is close to $100\%$. However, the success rate of the AIC decreases to about $94\%$ with a $6\%$ probability of overestimation.

\section{Summary and Outlook}
\label{sec:conclusion}
Concerning the performance of model selection, the results of our MC experiment can be summarized as follows. In alignment with similar studies for regression models, we can observe that the inconsistent AIC outperforms the other two IC when the MLE is applied to smaller samples. In contrast, the consistent IC (BIC and HQ) perform excellently for sufficiently large samples and we can observe a monotonic improvement in their success rate when increasing the sample size. In spite of the concerns presented above in Remark \ref{rem:defin-theor-prop-1} (iii), the numerical results show that the HQ should not be excluded as a well performing IC. More concerning, and not as commonly observed in previous studies, is a non-monotonic relation between sample size and success rate of model selection of the AIC.\\
Future research in this direction can be on the equally popular power law intensities and further model selection methods like the focused information criterion as well as model averaging.



\section*{Funding}

E. Scalas and M. Trinh were supported by the Strategic Development Fund of the University of Sussex.

\section*{Supplemental material}

Please contact either Enrico Scalas or Mailan Trinh.

\begin{tabbing}
\hspace*{4cm} \= \kill
\texttt{intensity.m} \> evaluates the intensity function of a Hawkes process\\
\texttt{hawkesThinning.m} \> simulates a Hawkes process with up to a specified time horizon\\
\texttt{hawkesThinning2.m} \> simulates a Hawkes process with up to a specified sample size\\
\texttt{empirAgg2.m} \> calculates average number of events based on simulated paths of a\\
\> Hawkes process (calls \texttt{hawkesThinning2.m})\\
\texttt{LogLik\_iter.m} \> evaluates the log-likelihood function of a Hawkes process for given\\
\> parameters and data\\
\texttt{constraints.m} \> parameter constrains passed on to the optimization algorithm \texttt{fmincon}\\
\texttt{fitting.m} \> maximizes log-likelihood function to obtain maximum likelihood\\
\> estimators using the MATLAB\tm routine \texttt{fmincon}\\
\> (calls \texttt{LogLik\_iter.m} and \texttt{constraints.m})\\
\texttt{IC.m} \> calculates the values of AIC, BIC and HQ
\end{tabbing}

\bibliographystyle{plain}
\bibliography{refHawkes}

\begin{thebibliography}{10}

\bibitem{Akaike_1973}
H.~Akaike.
\newblock Information theory and an extension of the maximum likelihood
  principle.
\newblock In {\em Second {I}nternational {S}ymposium on {I}nformation {T}heory
  ({T}sahkadsor, 1971)}, pages 267--281. Akad\'emiai Kiad\'o, Budapest, 1973.

\bibitem{Bacry_2012}
Emmanuel Bacry, Khalil Dayri, and Jean-Fran{\c{c}}ois Muzy.
\newblock Non-parametric kernel estimation for symmetric hawkes processes.
  application to high frequency financial data.
\newblock {\em The European Physical Journal B}, 85(5):1--12, 2012.

\bibitem{Bacry_2015}
Emmanuel Bacry, Iacopo Mastromatteo, and Jean-Fran\c{c}ois Muzy.
\newblock Hawkes processes in finance.
\newblock May 2015.

\bibitem{Bertram_2004}
William~K. Bertram.
\newblock An empirical investigation of australian stock exchange data.
\newblock {\em Physica A: Statistical Mechanics and its Applications},
  341:533--546, October 2004.

\bibitem{Bowsher_2007}
Clive~G. Bowsher.
\newblock Modelling security market events in continuous time: intensity based,
  multivariate point process models.
\newblock {\em J. Econometrics}, 141(2):876--912, 2007.

\bibitem{Bremaud_Massoulie_2001}
Pierre Br{\'e}maud and Laurent Massouli{\'e}.
\newblock Hawkes branching point processes without ancestors.
\newblock {\em J. Appl. Probab.}, 38(1):122--135, 2001.

\bibitem{Burnham_Anderson_2004}
Kenneth~P. Burnham and David~R. Anderson.
\newblock Multimodel inference: understanding {AIC} and {BIC} in model
  selection.
\newblock {\em Sociol. Methods Res.}, 33(2):261--304, 2004.

\bibitem{Claeskens_Hjort_2008}
Gerda Claeskens and Nils~Lid Hjort.
\newblock {\em Model selection and model averaging}.
\newblock Cambridge Series in Statistical and Probabilistic Mathematics.
  Cambridge University Press, Cambridge, 2008.

\bibitem{Daley_2003}
D.~J. Daley and D.~Vere-Jones.
\newblock {\em An introduction to the theory of point processes. {V}ol. {I}}.
\newblock Probability and its Applications (New York). Springer-Verlag, New
  York, second edition, 2003.
\newblock Elementary theory and methods.

\bibitem{Dassios_Zhao_2013}
Angelos Dassios and Hongbiao Zhao.
\newblock Exact simulation of {H}awkes process with exponentially decaying
  intensity.
\newblock {\em Electron. Commun. Probab.}, 18:no. 62, 13, 2013.

\bibitem{Engle_2000}
Robert~F Engle.
\newblock The econometrics of ultra-high-frequency data.
\newblock {\em Econometrica}, 68(1):1--22, 2000.

\bibitem{Engle_Russell_1997}
Robert~F Engle and Jeffrey~R Russell.
\newblock Forecasting the frequency of changes in quoted foreign exchange
  prices with the autoregressive conditional duration model.
\newblock {\em Journal of Empirical Finance}, 4(2):187--212, 1997.

\bibitem{Engle_Russell_1998}
Robert~F. Engle and Jeffrey~R. Russell.
\newblock Autoregressive conditional duration: a new model for irregularly
  spaced transaction data.
\newblock {\em Econometrica}, 66(5):1127--1162, 1998.

\bibitem{Filimonov_Sornette_2012}
Vladimir Filimonov and Didier Sornette.
\newblock Quantifying reflexivity in financial markets: Toward a prediction of
  flash crashes.
\newblock {\em Physical Review E}, 85(5):056108, 2012.

\bibitem{Freedman_1983}
David~A. Freedman.
\newblock A note on screening regression equations.
\newblock {\em Amer. Statist.}, 37(2):152--155, 1983.

\bibitem{Hannan_1980}
E.~J. Hannan.
\newblock The estimation of the order of an {ARMA} process.
\newblock {\em Ann. Statist.}, 8(5):1071--1081, 1980.

\bibitem{Hannan_1979}
E.~J. Hannan and B.~G. Quinn.
\newblock The determination of the order of an autoregression.
\newblock {\em J. Roy. Statist. Soc. Ser. B}, 41(2):190--195, 1979.

\bibitem{Hardiman_2013}
Stephen~J Hardiman, Nicolas Bercot, and Jean-Philippe Bouchaud.
\newblock Critical reflexivity in financial markets: a hawkes process analysis.
\newblock {\em The European Physical Journal B}, 86(442), 2013.

\bibitem{Hautsch_2012}
Nikolaus Hautsch.
\newblock {\em Econometrics of financial high-frequency data}.
\newblock Springer, Heidelberg, 2012.

\bibitem{Hawkes_1973}
A.G. Hawkes and L.~Adamopoulos.
\newblock Cluster models for earthquakes-regional comparisons.
\newblock {\em Bull. Int. Statist. Inst}, 45(3):454--461, 1973.

\bibitem{Hawkes_1971b}
Alan~G. Hawkes.
\newblock Point spectra of some mutually exciting point processes.
\newblock {\em J. Roy. Statist. Soc. Ser. B}, 33:438--443, 1971.

\bibitem{Hawkes_1971a}
Alan~G. Hawkes.
\newblock Spectra of some self-exciting and mutually exciting point processes.
\newblock {\em Biometrika}, 58:83--90, 1971.

\bibitem{Hurvich_Tsai_1989}
Clifford~M. Hurvich and Chih-Ling Tsai.
\newblock Regression and time series model selection in small samples.
\newblock {\em Biometrika}, 76(2):297--307, 1989.

\bibitem{Hurvich_Tsai_1990}
Clifford~M Hurvich and Chih—Ling Tsai.
\newblock The impact of model selection on inference in linear regression.
\newblock {\em The American Statistician}, 44(3):214--217, 1990.

\bibitem{Javed_Mantalos_2013}
Farrukh Javed and Panagiotis Mantalos.
\newblock G{ARCH}-type models and performance of information criteria.
\newblock {\em Comm. Statist. Simulation Comput.}, 42(8):1917--1933, 2013.

\bibitem{Lallouache_Challet_2016}
Mehdi Lallouache and Damien Challet.
\newblock The limits of statistical significance of {H}awkes processes fitted
  to financial data.
\newblock {\em Quant. Finance}, 16(1):1--11, 2016.

\bibitem{Lewis_Shedler_1979}
P.~A.~W. Lewis and G.~S. Shedler.
\newblock Simulation of nonhomogeneous {P}oisson processes by thinning.
\newblock {\em Naval Res. Logist. Quart.}, 26(3):403--413, 1979.

\bibitem{Meyer_1971}
P.~A. Meyer.
\newblock D\'emonstration simplifi\'ee d'un th\'eor\`eme de {K}night.
\newblock pages 191--195. Lecture Notes in Math., Vol. 191, 1971.

\bibitem{Moller_Rasmussen_2005}
Jesper M{\o}ller and Jakob~G. Rasmussen.
\newblock Perfect simulation of {H}awkes processes.
\newblock {\em Adv. in Appl. Probab.}, 37(3):629--646, 2005.

\bibitem{Moller_Rasmussen_2006}
Jesper M{\o}ller and Jakob~G. Rasmussen.
\newblock Approximate simulation of {H}awkes processes.
\newblock {\em Methodol. Comput. Appl. Probab.}, 8(1):53--64, 2006.

\bibitem{Note1}
By ``small samples'' we refer to the situation that the sample size is not
  sufficiently large enough for the asymptotic consistency results to hold, but
  large enough such that effects similar to the paradox discussed by \cite
  {Freedman_1983} can be safely excluded.

\bibitem{Ogata_1978}
Yosihiko Ogata.
\newblock The asymptotic behaviour of maximum likelihood estimators for
  stationary point processes.
\newblock {\em Ann. Inst. Statist. Math.}, 30(2):243--261, 1978.

\bibitem{Ogata_1981}
Yosihiko Ogata.
\newblock On {L}ewis' simulation method for point processes.
\newblock {\em Information Theory, IEEE Transactions on}, 27(1):23--31, 1981.

\bibitem{Ogata_1988}
Yosihiko Ogata.
\newblock Statistical models for earthquake occurrences and residual analysis
  for point processes.
\newblock {\em Journal of the American Statistical association}, 83(401):9--27,
  1988.

\bibitem{Ozaki_1979}
T.~Ozaki.
\newblock Maximum likelihood estimation of {H}awkes' self-exciting point
  processes.
\newblock {\em Ann. Inst. Statist. Math.}, 31(1):145--155, 1979.

\bibitem{Rambaldi_2015}
Marcello Rambaldi, Paris Pennesi, and Fabrizio Lillo.
\newblock Modeling foreign exchange market activity around macroeconomic news:
  Hawkes-process approach.
\newblock {\em Physical Review E}, 91(1):012819, 2015.

\bibitem{Rasmussen_2013}
Jakob~Gulddahl Rasmussen.
\newblock Bayesian inference for {H}awkes processes.
\newblock {\em Methodol. Comput. Appl. Probab.}, 15(3):623--642, 2013.

\bibitem{Schwarz_1978}
Gideon Schwarz.
\newblock Estimating the dimension of a model.
\newblock {\em Ann. Statist.}, 6(2):461--464, 1978.

\bibitem{Shibata_1976}
Ritei Shibata.
\newblock Selection of the order of an autoregressive model by {A}kaike's
  information criterion.
\newblock {\em Biometrika}, 63(1):117--126, 1976.

\bibitem{Sin_White_1996}
Chor-Yiu Sin and Halbert White.
\newblock Information criteria for selecting possibly misspecified parametric
  models.
\newblock {\em J. Econometrics}, 71(1-2):207--225, 1996.

\bibitem{Veen_Schoenberg_2008}
Alejandro Veen and Frederic~P. Schoenberg.
\newblock Estimation of space-time branching process models in seismology using
  an {EM}-type algorithm.
\newblock {\em J. Amer. Statist. Assoc.}, 103(482):614--624, 2008.

\bibitem{Zhang_1993}
Ping Zhang.
\newblock On the convergence rate of model selection criteria.
\newblock {\em Comm. Statist. Theory Methods}, 22(10):2765--2775, 1993.

\end{thebibliography}

\subsection{Appendices}

\begin{table}[htbp]
\center
\caption{Simulation parameters (Parameter Set 1)}
\begin{tabular}{|l|r|r|l|l|r|l|l|}
\hline
 & \multicolumn{1}{c|}{$\mu$} & \multicolumn{1}{c|}{$\alpha_1$} & \multicolumn{1}{c|}{$\alpha_2$} & \multicolumn{1}{c|}{$\alpha_3$} & \multicolumn{1}{c|}{$\beta_1$} & \multicolumn{1}{c|}{$\beta_2$} & \multicolumn{1}{c|}{$\beta_3$} \\ \hline
P=1 & 0.5 & 9 & -- & -- & 10 & -- & -- \\ \hline
P=2 & 0.5 & 0.00066 & \multicolumn{1}{r|}{100} & -- & 0.001 & \multicolumn{1}{r|}{300} & -- \\ \hline
P=3 & 0.5 & 0.00033 & \multicolumn{1}{r|}{3.3} & \multicolumn{1}{r|}{100} & 0.001 & \multicolumn{1}{r|}{10} & \multicolumn{1}{r|}{300} \\ \hline
\end{tabular}
\label{hawkesSim}
\end{table} 

\begin{table}[htbp]
\caption{Parameter set taken from \cite{Lallouache_Challet_2016} (Parameter Set 2)}
\begin{center}
\begin{tabular}{|l|l|l|l|l|}
\hline
\multicolumn{1}{|c|}{$\mu$} & \multicolumn{1}{c|}{$\alpha_1$} & \multicolumn{1}{c|}{$\alpha_2$} & \multicolumn{1}{c|}{$\beta_1$} & \multicolumn{1}{c|}{$\beta_2$} \\ \hline
\multicolumn{1}{|r|}{0.05}&\multicolumn{1}{r|}{0.01761905} & \multicolumn{1}{r|}{0.28} & \multicolumn{1}{r|}{0.04761905} & \multicolumn{1}{r|}{0.6666667} \\ \hline
\end{tabular}
\end{center}
\label{param_lallouache}
\end{table}

\begin{table}[htbp]
\small
\caption{Absolute RMSE values for MLE of the exponential Hawkes models of order $P\in\{1,2,3\}$ using Parameter Set 1 with varying time horizons $T\in \{500, 1000, 2000, 5000\}$. The order of the model which was used for simulation coincides with the model used for fitting. Thus, the true parameter values are known and the RMSE is expected to decrease as the MLE improves.}
\begin{center}
\begin{tabular}{|l|l|r|r|r|l|r|r|l|r|}
\hline
 &  & \multicolumn{1}{c|}{$\mu$} & \multicolumn{1}{c|}{$\alpha_1$} & \multicolumn{1}{c|}{$\alpha_2$}& \multicolumn{1}{c|}{$\alpha_3$} & \multicolumn{1}{c|}{$\beta_1$} & \multicolumn{1}{c|}{$\beta_2$} & \multicolumn{1}{c|}{$\beta_3$} &\multicolumn{1}{c|}{\parbox[t]{2cm}{Average \\sample size}}\\ \hline\hline
P=1 & T=500 & 0.039664 & 0.42256 & \multicolumn{1}{l|}{--} &\multicolumn{1}{l|}{--}& 0.44731 & \multicolumn{1}{l|}{--} & -- & 2483\\ \hline
 & T=1000 & 0.02763 & 0.32473 & \multicolumn{1}{l|}{--} &\multicolumn{1}{l|}{--} & 0.32928 & \multicolumn{1}{l|}{-} & -- & 5019\\ \hline
 & T=2000 & 0.018738 & 0.21756 & \multicolumn{1}{l|}{--} &\multicolumn{1}{l|}{--} & 0.22197 & \multicolumn{1}{l|}{-} & -- & 9977\\ \hline
 & T=5000 & 0.011276 & 0.13846 & \multicolumn{1}{l|}{--} &\multicolumn{1}{l|}{--} & 0.14544 & \multicolumn{1}{l|}{-} & -- & 24962\\ \hline\hline
P=2 & T=500 & 0.071796 & 6.0214 & 12.399&-- & 37.322 & 44.92 & -- & 470\\ \hline
 & T=1000 & 0.061258 & 0.00085434 & 7.9865&-- & 0.0023803 & 18.956 & -- & 1121\\ \hline
 & T=2000 & 0.049989 & 0.00020347 & 4.7732&-- & 0.00045077 & 11.415 & -- & 2977\\ \hline
 & T=5000 & 0.042938 & 0.00010551 & 2.3918&-- & 0.00018339 & 5.9634 & -- & 12883\\ \hline\hline
P=3 & T=500 & 0.07713 & 0.3232 & 1.3408&\multicolumn{1}{r|}{9.8602} & 1.5085 & 9.7124 & \multicolumn{1}{r|}{30.678} & 929\\ \hline
 & T=1000 & 0.061804 & 0.00036118 & 0.29469&\multicolumn{1}{r|}{6.2401} & 0.0021189 & 0.76784 & \multicolumn{1}{r|}{18.334} & 2207\\ \hline
 & T=2000 & 0.051527 & 0.0001279 & 0.19655&\multicolumn{1}{r|}{3.8663} & 0.00058096 & 0.49288 & \multicolumn{1}{r|}{11.77} & 5840\\ \hline
 & T=5000 & 0.045562 & 0.000055755 & 0.10766&\multicolumn{1}{r|}{1.8741} & 0.00019317 & 0.27921 & \multicolumn{1}{r|}{5.6723} & 25017\\ \hline
\end{tabular}
\end{center}
\label{rmse}
\end{table}

\begin{table}[htbp]
\caption{Relative RMSE values for MLE of the exponential Hawkes models of order $P\in\{1,2,3\}$ using Parameter Set 1 with varying time horizons $T\in \{500, 1000, 2000, 5000\}$. The order of the model which was used for simulation coincides with the model used for fitting. Thus, the true parameter values are known and the RMSE is expected to decrease as the MLE improves. The values are given in percent.}
\begin{center}
\begin{tabular}{|l|l|r|r|r|l|r|r|l|r|}
\hline
 &  & \multicolumn{1}{c|}{$\mu$} & \multicolumn{1}{c|}{$\alpha_1$} & \multicolumn{1}{c|}{$\alpha_2$}&\multicolumn{1}{c|}{$\alpha_3$} & \multicolumn{1}{c|}{$\beta_1$} & \multicolumn{1}{c|}{$\beta_2$} & \multicolumn{1}{c|}{$\beta_3$} & \multicolumn{1}{r|}{\parbox[t]{2cm}{Average \\sample size}}\\ \hline\hline
P=1 & T=500 & 7.9328 & 4.6951 & \multicolumn{1}{l|}{--}&-- & 4.4731 & \multicolumn{1}{l|}{--} & -- & 2483\\ \hline
 & T=1000 & 5.526 & 3.6082 & \multicolumn{1}{l|}{--}&-- & 3.2928 & \multicolumn{1}{l|}{--} & -- &5019\\ \hline
 & T=2000 & 3.7475 & 2.4173 & \multicolumn{1}{l|}{--}&-- & 2.2197 & \multicolumn{1}{l|}{--} & -- &9977\\ \hline
 & T=5000 & 2.2551 & 1.5384 & \multicolumn{1}{l|}{--}&-- & 1.4544 & \multicolumn{1}{l|}{--} & -- &24962\\ \hline\hline
P=2 & T=500 & 14.359 & 912330 & 12.399&-- & 3732200 & 14.973 & -- &470\\ \hline
 & T=1000 & 12.252 & 129.45 & 7.9865&-- & 238.03 & 6.3186 & -- &1121\\ \hline
 & T=2000 & 9.9978 & 30.829 & 4.7732&-- & 45.077 & 3.805 & -- &2977\\ \hline
 & T=5000 & 8.5877 & 15.986 & 2.3918&-- & 18.339 & 1.9878 & -- &12883\\ \hline\hline
P=3 & T=500 & 15.426 & 97939 & 40.63&\multicolumn{1}{r|}{9.8602} & 150850 & 97.124 & \multicolumn{1}{r|}{10.226} &929\\ \hline
 & T=1000 & 12.361 & 109.45 & 8.9301&\multicolumn{1}{r|}{6.2401} & 211.89 & 7.6784 & \multicolumn{1}{r|}{6.1113} &2207\\ \hline
 & T=2000 & 10.305 & 38.758 & 5.956&\multicolumn{1}{r|}{3.8663} & 58.096 & 4.9288 & \multicolumn{1}{r|}{3.9233} &5840\\ \hline
 & T=5000 & 9.1124 & 16.895 & 3.2624&\multicolumn{1}{r|}{1.8741} & 19.317 & 2.7921 & \multicolumn{1}{r|}{1.8908} &25017\\ \hline
\end{tabular}
\end{center}
\label{rmseRel}
\end{table}

\begin{table}[htbp]
\caption{Model selection for simulated data of an exponential Hawkes model of order P=1 using Parameter Set 1 with varying time horizons $T\in \{500, 1000, 2000, 5000\}$. The numbers indicate how often the model order $P\in\{1,2,3\}$ is selected among the $1000$ samples and are given in percent. Bold numbers show which model was selected most often.}
\begin{center}
\begin{tabular}{|l|l|r|r|r|r|}
\hline
 & Time horizon&\multicolumn{1}{l|}{P=1} & \multicolumn{1}{l|}{P=2} & \multicolumn{1}{l|}{P=3} & \multicolumn{1}{l|}{\parbox[t]{2cm}{Average \\sample size}} \\ \hline\hline
AIC & T=500&\textbf{92.8} & 6.9 & 0.3 & 2483 \\ \hline
 & T=1000&\textbf{91.6} & 7.9 & 0.5 & 5019 \\ \hline
 & T=2000&\textbf{92.1} & 7.6 & 0.3 & 9977 \\ \hline
 & T=5000&\textbf{93.7} & 6.1 & 0.2 & 24962 \\ \hline\hline
BIC & T=500&\textbf{99.8} & 0.2 & 0 & 2483 \\ \hline
 & T=1000&\textbf{100} & 0 & 0 & 5019 \\ \hline
 & T=2000&\textbf{100} & 0 & 0 & 9977 \\ \hline
 & T=5000&\textbf{100} & 0 & 0 & 24962 \\ \hline\hline
HQ & T=500&\textbf{98.9} & 1.1 & 0 & 2483 \\ \hline
 & T=1000&\textbf{98.6} & 1.2 & 0.2 & 5019 \\ \hline
 & T=2000&\textbf{99.2} & 0.8 & 0 & 9977 \\ \hline
 & T=5000&\textbf{99.7} & 0.3 & 0 & 24962 \\ \hline
\end{tabular}
\end{center}
\label{modelSelection01}
\end{table}

\begin{table}[htbp]
\caption{Model selection for simulated data of an exponential Hawkes model of order P=2 using Parameter Set 1 with varying time horizons $T\in \{500, 1000, 2000, 5000\}$. The numbers indicate how often the model order $P\in\{1,2,3\}$ is selected among the $1000$ samples and are given in percent. Bold numbers show which model was selected most often.}
\begin{center}
\begin{tabular}{|l|l|r|r|r|r|}
\hline
 & Time horizon&\multicolumn{1}{l|}{P=1} & \multicolumn{1}{l|}{P=2} & \multicolumn{1}{l|}{P=3} & \multicolumn{1}{l|}{\parbox[t]{2cm}{Average \\sample size}} \\ \hline\hline
AIC & T=500&48.2 & \textbf{50.3} & 1.5 & 470 \\ \hline
 & T=1000&0.2 & \textbf{99} & 0.8 & 1121 \\ \hline
 & T=2000&0 & \textbf{96.9} & 3.1 & 2977 \\ \hline
 & T=5000&0 & \textbf{93.7} & 6.3 & 12883 \\ \hline\hline
BIC & T=500&\textbf{94.5} & 5.4 & 0.1 & 470 \\ \hline
 & T=1000&10.3 & \textbf{89.7} & 0 & 1121 \\ \hline
 & T=2000&0 & \textbf{100} & 0 & 2977 \\ \hline
 & T=5000&0 & \textbf{100} & 0 & 12883 \\ \hline\hline
HQ & T=500&\textbf{76.9} & 22.9 & 0.2 & 470 \\ \hline
 & T=1000&2.1 & \textbf{97.8} & 0.1 & 1121 \\ \hline
 & T=2000&0 & \textbf{99.4} & 0.6 & 2977 \\ \hline
 & T=5000&0 & \textbf{99.6} & 0.4 & 12883 \\ \hline
\end{tabular}
\end{center}
\label{modelSelection02}
\end{table}

\begin{table}[htbp]
\caption{Model selection for simulated data of an exponential Hawkes model of order P=3 using Parameter Set 1 with varying time horizons $T\in \{500, 1000, 2000, 5000\}$. The numbers indicate how often the model order $P\in\{1,2,3\}$ is selected among the $1000$ samples and are given in percent. Bold numbers show which model was selected most often.}
\begin{center}
\begin{tabular}{|l|l|r|r|r|r|}
\hline
 & Time horizon&\multicolumn{1}{l|}{P=1} & \multicolumn{1}{l|}{P=2} & \multicolumn{1}{l|}{P=3} & \multicolumn{1}{l|}{\parbox[t]{2cm}{Average \\sample size}} \\ \hline\hline
AIC & T=500&0 & \textbf{53.7} & 46.3 & 929 \\ \hline
 & T=1000&0 & 0.2 & \textbf{99.8} & 2207 \\ \hline
 & T=2000&0 & 0 & \textbf{100} & 5840 \\ \hline
 & T=5000&0 & 0 & \textbf{100} & 25017 \\ \hline\hline
BIC & T=500&0 & \textbf{96.5} & 3.5 & 929 \\ \hline
 & T=1000&0 & 25 & \textbf{75} & 2207 \\ \hline
 & T=2000&0 & 0 & \textbf{100} & 5840 \\ \hline
 & T=5000&0 & 0 & \textbf{100} & 25017 \\ \hline\hline
HQ & T=500&0 & \textbf{81.6} & 18.4 & 929 \\ \hline
 & T=1000&0 & 4.7 & \textbf{95.3} & 2207 \\ \hline
 & T=2000&0 & 0 & \textbf{100} & 5840 \\ \hline
 & T=5000&0 & 0 & \textbf{100} & 25017 \\ \hline
\end{tabular}
\end{center}
\label{modelSelection03}
\end{table}

\begin{table}[htbp]
\caption{Absolute RMSE values for MLE of the exponential Hawkes models of order $P=2$ using Parameter Set 2 with varying time horizons $T\in \{600, 900, 1800, 3600, 7200, 21600\}$. The order of the model which was used for simulation coincides with the model used for fitting. Thus, the true parameter values are known and the RMSE is expected to decrease as the MLE improves.}
\begin{center}
\begin{tabular}{|l|l|r|r|r|r|r|r|}
\hline
  & \multicolumn{1}{c|}{$\mu$}&\multicolumn{1}{c|}{$\alpha_1$} & \multicolumn{1}{c|}{$\alpha_2$} & \multicolumn{1}{c|}{$\beta_1$} & \multicolumn{1}{c|}{$\beta_2$} & \multicolumn{1}{l|}{\parbox[t]{2cm}{Average \\sample size}}\\ \hline\hline
T=600 & 0.030176&0.076208 & 12489000 & 0.17692 & 27063000000 & 135\\ \hline
T=900 & 0.025411&0.052971 & 1704000 & 0.11054 & 15861000000 & 205\\ \hline
T=1800 & 0.016031&0.023916 & 0.078183 & 0.049322 & 4.0458 & 417\\ \hline
T=3600 & 0.010572&0.0095225 & 0.039458 & 0.020001 & 0.15516 & 853\\ \hline
T=7200 & 0.0070848&0.0055058 & 0.025844 & 0.011919 & 0.088505 & 1708\\ \hline
T=21600 & 0.0039548&0.0030036 & 0.014737 & 0.006448 & 0.051022 & 5144\\ \hline
\end{tabular}
\end{center}
\label{rmselal}
\end{table}

\begin{table}[htbp]
\caption{Relative RMSE values for MLE of the exponential Hawkes models of order $P=2$ using Parameter Set 2 with varying time horizons $T\in \{600, 900, 1800, 3600, 7200, 21600\}$. The order of the model which was used for simulation coincides with the model used for fitting. Thus, the true parameter values are known and the RMSE is expected to decrease as the MLE improves. The values are given in percent.}
\begin{center}
\begin{tabular}{|l|l|r|r|r|r|r|r|}
\hline
& \multicolumn{1}{c|}{$\mu$}&\multicolumn{1}{c|}{$\alpha_1$} & \multicolumn{1}{c|}{$\alpha_2$} & \multicolumn{1}{c|}{$\beta_1$} & \multicolumn{1}{c|}{$\beta_2$} & \multicolumn{1}{l|}{\parbox[t]{2cm}{Average \\sample size}}\\ \hline\hline
 T=600 & 60.353&432.53 & 4460300000 & 371.53 & 4059400000000 & 135\\ \hline
 T=900 & 50.822&300.64 & 608590000 & 232.14 & 2379200000000 & 205\\ \hline
 T=1800 & 32.061&135.74 & 27.923 & 103.58 & 606.87 & 417\\ \hline
 T=3600 & 21.144&54.047 & 14.092 & 42.003 & 23.275 & 853\\ \hline
 T=7200 & 14.17&31.249 & 9.2299 & 25.03 & 13.276 & 1708\\ \hline
 T=21600 & 7.9096&17.047 & 5.263 & 13.541 & 7.6533 & 5144\\ \hline
\end{tabular}
\end{center}
\label{rmseRellal}
\end{table}

\begin{table}[htbp]
\caption{Model selection for simulated data of an exponential Hawkes model of order P=2 using Parameter Set 2 with varying time horizons $T\in \{600, 900, 1800, 3600, 7200, 21600\}$. The numbers indicate how often the model order $P\in\{1,2,3\}$ is selected among the $1000$ samples and are given in percent. Bold numbers show which model was selected most often.}
\begin{center}
\begin{tabular}{|l|l|r|r|r|r|}
\hline
 & Time horizon &\multicolumn{1}{l|}{P=1} & \multicolumn{1}{l|}{P=2} & \multicolumn{1}{l|}{P=3} & \multicolumn{1}{l|}{\parbox[t]{2cm}{Average \\sample size}} \\ \hline\hline
AICc/AIC & T=600&\textbf{51.4} & 48 & 0.6 & 135 \\ \hline
 & T=900&35.1 & \textbf{63.8} & 1.1 & 205 \\ \hline
 & T=1800&6.7 & \textbf{90.2} & 3.1 & 417 \\ \hline
 & T=3600&0 & \textbf{96.9} & 3.1 & 853 \\ \hline
 & T=7200&0 & \textbf{94.7} & 5.3 & 1708 \\ \hline
 & T=21600&0 & \textbf{94.1} & 5.9 & 5144 \\ \hline\hline
AIC & T=600&49.3 & \textbf{50.1} & 0.6 & 135 \\ \hline
 & T=900&34.8 & \textbf{64.1} & 1.1 & 205 \\ \hline
 & T=1800&6.7 & \textbf{90.2} & 3.1 & 417 \\ \hline
 & T=3600&0 & \textbf{96.9} & 3.1 & 853 \\ \hline
 & T=7200&0 & \textbf{94.7} & 5.3 & 1708 \\ \hline
 & T=21600&0 & \textbf{94.1} & 5.9 & 5144 \\ \hline\hline
BIC & T=600&\textbf{86.5} & 13.5 & 0 & 135 \\ \hline
 & T=900&\textbf{79.8} & 20.2 & 0 & 205 \\ \hline
 & T=1800&42.7 & \textbf{57.2} & 0.1 & 417 \\ \hline
 & T=3600&5 & \textbf{95} & 0 & 853 \\ \hline
 & T=7200&0.1 & \textbf{99.9} & 0 & 1708 \\ \hline
 & T=21600&0 & \textbf{100} & 0 & 5144 \\ \hline\hline
HQ & T=600&\textbf{69} & 30.8 & 0.2 & 135 \\ \hline
 & T=900&\textbf{55.6} & 44.4 & 0 & 205 \\ \hline
 & T=1800&17.5 & \textbf{81.8} & 0.7 & 417 \\ \hline
 & T=3600&1 & \textbf{98.7} & 0.3 & 853 \\ \hline
 & T=7200&0 & \textbf{98.9} & 1.1 & 1708 \\ \hline
 & T=21600&0 & \textbf{99.2} & 0.8 & 5144 \\ \hline
\end{tabular}
\end{center}
\label{modelSelectionlallouache}
\end{table}

\end{document}